# An Update on a Progressively Expanded Database for Automated Lung Sound Analysis

Fu-Shun Hsu[1,2,3], Shang-Ran Huang[3], Chien-Wen Huang[4], Yuan-Ren Cheng[3,5,6], Chun-Chieh Chen[4], Jack Hsiao[7], Chung-Wei Chen[2], and Feipei Lai[1,*]

[1] Graduate Institute of Biomedical Electronics and Bioinformatics, National Taiwan University, Taipei, Taiwan

[2] Department of Critical Care Medicine, Far Eastern Memorial Hospital, New Taipei, Taiwan

[3] Heroic Faith Medical Science Co., Ltd., Taipei, Taiwan

[4] Avalanche Computing Inc., Taipei, Taiwan

[5] Department of Life Science, College of Life Science, National Taiwan University, Taipei, Taiwan

[6] Institute of Biomedical Sciences, Academia Sinica, Taipei, Taiwan

[7] HCC Healthcare Group, New Taipei, Taiwan

[*] Corresponding author

E-mail: flai@csie.ntu.edu.tw

**Acknowledgments**

Taiwan's Raising Children Medical Foundation sponsored the lung sound collection activity. Heroic Faith Medical Science Co., Ltd. sponsored the data labeling and deep learning model training. This work was supported by Taiwan's Ministry of Science and Technology under grant number MOST 109-EC-17-A-22-I3-0009. The authors thank Taiwan's National Center for High-Performance Computing for providing the computing resources. The authors also thank the employees of Heroic Faith Medical Science Co., Ltd., who contributed to the establishment of the HF_Lung_V2 database. This manuscript was edited by Wallace Academic Editing.

# Abstract

**Purpose** We previously established an open-access lung sound database, HF_Lung_V1, and developed deep learning models for inhalation, exhalation, continuous adventitious sound (CAS), and discontinuous adventitious sound (DAS) detection. The amount of data used for training contributes to model accuracy. Herein, we collected larger quantities of data to further improve model performance. Moreover, the issues of noisy labels and sound overlapping were explored.

**Methods** HF_Lung_V1 was expanded to HF_Lung_V2 with a 1.45× increase in the number of audio files. Convolutional neural network–bidirectional gated recurrent unit network models were trained separately using the HF_Lung_V1 (V1_Train) and HF_Lung_V2 (V2_Train) training sets and then tested using the HF_Lung_V1 (V1_Test) and HF_Lung_V2 (V2_Test) test sets, respectively. Segment and event detection performance was evaluated using the F1 scores. Label quality was assessed. Moreover, the overlap ratios between inhalation, exhalation, CAS, and DAS labels were computed.

**Results** The model trained using V2_Train exhibited improved F1 scores in inhalation, exhalation, and CAS detection on both V1_Test and V2_Test but not in DAS detection. Poor CAS detection was attributed to the quality of CAS labels. DAS detection was strongly influenced by the overlapping of DAS labels with inhalation and exhalation labels.

**Conclusion** Collecting greater quantities of lung sound data is vital for developing more accurate lung sound analysis models. To build real ground-truth labels, the labels must be reworked; this process is ongoing. Furthermore, a method for addressing the sound overlapping problem in DAS detection must be formulated.

**Keywords:** Auscultation, Convolutional neural network, Deep learning, Gated recurrent unit, Lung sound

# 1 Introduction

Respiration is an essential vital sign. Changes in the frequency or intensity of respiratory sounds and the manifestation of continuous adventitious sounds (CASs), such as wheezes, stridor, and rhonchi, and discontinuous adventitious sounds (DASs), such as crackles and pleural friction rubs, are associated with pulmonary disorders [1, 2]. Detection of adventitious sounds during respiratory sound auscultation can help health-care professionals make clinical decisions. Methods of automated adventitious respiratory sound analysis as of 2020 have been comprehensively reviewed [3, 4]. Identification of the breathing phase (inhalation and exhalation) during respiratory sound auscultation can help clinicians estimate the respiratory rate and diagnose apnea [5, 6].

Most previously reported automated respiratory sound analysis algorithms focused only on sound type classification or normal–abnormal classification but not on sound event detection [7]. Messner et al. [8] reported F1 scores of approximately 86% and approximately 72% for breath phase detection and crackle detection, respectively, using a bidirectional gated recurrent unit (BiGRU) network. Jácome et al. [9] employed a faster region-based convolutional neural network (CNN) framework, obtaining a sensitivity and specificity of 97.5% and 85%, respectively, in inspiratory phase detection, and a sensitivity and specificity of 95.5% and 82.5%, respectively, in expiratory phase detection. These studies were all limited by small data volumes.

In a previous study, we developed the world's largest open-access lung sound database, HF_Lung_V1 (https://gitlab.com/techsupportHF/HF_Lung_V1), and used several variants of recurrent neural networks (RNNs) to benchmark HF_Lung_V1 [7]. Results indicated the potential of using deep learning for automated inhalation, exhalation, CAS, and DAS detection. Because the performance of deep learning models is positively related to the size of the training set [10], we present an update herein on expanding HF_Lung_V1 to HF_Lung_V2 by using more lung sound files and labels. We determined whether the detection model trained using HF_Lung_V2 exhibited performance improvement with an increase in the data size. Moreover, given the potential presence of noisy labels in the database, label quality was reviewed. In addition, the extent of sound overlapping was investigated.

## 2 Materials and Methods

**2.1 Participants and Lung Sound Recording**

The lung sound database HF_Lung_V2 is an incremental expansion of a previously developed database, HF_Lung_V1. To build HF_Lung_V2, additional lung sounds were collected between October 2019 and December 2019. Lung sounds were recorded using a commercially available electronic stethoscope, Littmann 3200 (3M, Saint Paul, Minnesota, USA), and a custom multichannel sound recording device, HF-Type-1 (Heroic Faith Medical Science, Taipei, Taiwan) [7].

The Littmann 3200 stethoscope can only record breath sounds at a single location at a time. Therefore, each round of breath sound recordings was sequentially conducted at eight locations, namely the second intercostal space (ICS) in the right and left midclavicular lines (MCLs), the fifth ICS in the right and left MCLs, the fourth ICS in the right and left midaxillary lines (MALs), and the 10th ICS in the right and left MALs. In addition, the HF-Type-1 was used to simultaneously record lung sounds from six locations (the same as the aforementioned locations except for the fourth ICS in the right and left MALs) by multichannel acoustic recordings. A complete round of recording with the HF-Type-1 consisted of the acquisition of continuous signals from each of these six locations over 30 min. The detailed recording protocol is provided in our previous study [7].

We collected lung sounds from 10 residents of a respiratory care ward (RCW) or a respiratory care center (RCC) who were under long-term mechanical ventilation support by using the Littmann 3200 for 4-5 rounds and by using HF-Type-1 for 3-4 rounds. The lung sounds of another 32 inpatients with adventitious sounds at Far Eastern Memorial Hospital were collected using the Littmann 3200 alone for 1–3 rounds. All the participants were Taiwanese and aged older than 20 years. The sampling rate of the recording devices was 4000 Hz and the bit depth was 16 bits.

**2.2 Data Preparation and Labeling**

The length of audio files recorded using the Littmann 3200 was originally set as 15.8 s; as such, the terminal 0.8-s segment was deleted to obtain 15-s audio files. As for the audio files recorded using the HF-Type-1, the first 15-s segment of every 2-min recording was truncated for subsequent analysis. Two board-certified respiratory therapists with 8 and 4 years of clinical experience respectively and one board-certified nurse with 13 years of clinical experience performed the labeling. Each lung sound file was labeled by only one individual, and regular consensus meetings were held to ensure that all labelers employed the same labeling criteria. A self-developed software was used to label inhalations, exhalations, wheezes, stridor, rhonchi, and crackles [11]. Labels of wheezes (W), stridor (S), and rhonchi (R) were combined to obtain CAS labels (C), whereas DAS labels (D) contained all types of crackles without pleural friction rubs.

## 2.3 Dataset Arrangement

The acoustic patterns of lung sounds collected from one individual at different auscultation locations or those recorded at short time intervals might exhibit numerous similarities. Therefore, the audio files from the same individual were randomly distributed to either the training set or the test set to avoid potential data leakage. Based on the number of recordings, the training set–test set ratio was maintained at approximately 4:1.

## 2.4 Training Process

The CNN–BiGRU model presented in Fig. 1 outperformed all benchmark models in our previous study [9]. Therefore, the same CNN–BiGRU model was used to benchmark HF_Lung_V2 in the present study.

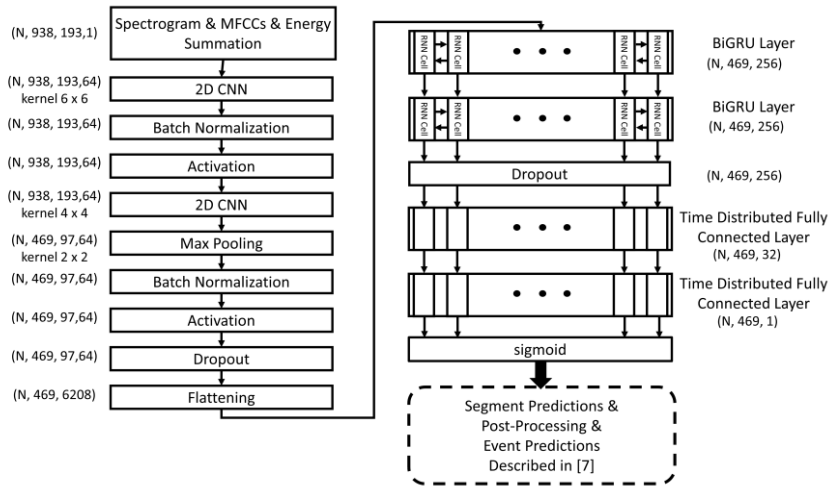

**Fig. 1** Architecture of the CNN–BiGRU model

As shown in Fig. 2, the pipeline of preprocessing, deep learning processing, and postprocessing was the same as that described in our previous study [7]. First, the obtained signal was processed using a high-pass filter with a cutoff frequency of 80 Hz. Next, the spectrogram, Mel-frequency cepstral coefficients (MFCCs) [12], and energy summation were calculated from the filtered signal. Using the short-time Fourier transform with a Hanning window size of 256, a hop length of 64, and no zero padding, the spectrogram was generated. The MFCCs comprised 20 static coefficients, 20 delta coefficients, and 20 acceleration coefficients. Energy summation is the summed energy in four frequency bands: 0–250, 250–500, 500–1000, and 0–2000 Hz. More details can be found in our previous study [7].

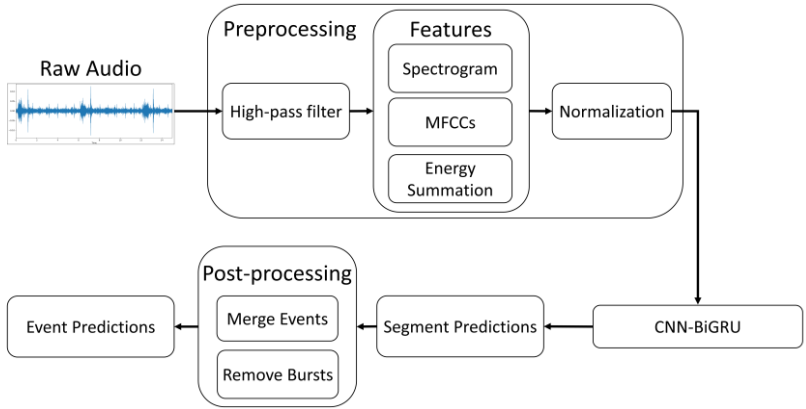

**Fig. 2** Pipeline of preprocessing, deep learning processing, and postprocessing

We used only audio files containing CASs and DASs to train and test the corresponding detection models. Fivefold cross-validation was conducted during the training process; to obtain more stable results, this procedure was repeated thrice. The performance measurements of 15 models obtained from the repeated fivefold cross-validation were averaged and reported.

**2.5 Performance Evaluation**

Fig. 3 illustrates how we evaluated the segment and event detection performance. First, the ground-truth event labels (red horizontal bars in Fig. 3a) were used to generate the ground-truth time segments (red vertical bars in Fig. 3b). Next, by comparing the results of segment prediction (blue vertical bars in Fig. 3c) with the ground-truth time segments (Fig. 3b), we defined true-positive (TP; orange vertical bars in Fig. 3e), true-negative (TN; green vertical bars in Fig. 3e), false-positive (FP; black vertical bars in Fig. 3e), and false-negative (FN; yellow vertical bars in Fig. 3e) time segments.

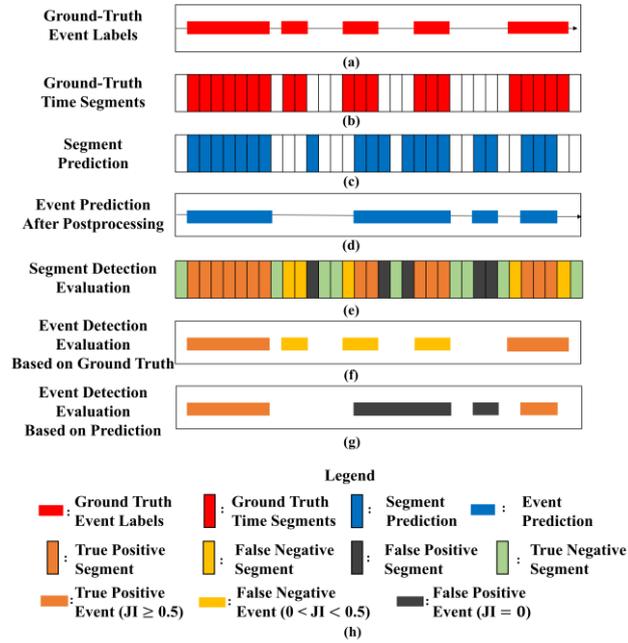

**Fig. 3** Illustration of segment and event detection evaluation: (a) Ground-truth event labels, (b) ground-truth time segments, (c) segment prediction, (d) event prediction after postprocessing, (e) segment detection evaluation, (f) event evaluation based on ground-truth event labels, (g) event detection evaluation based on prediction, and (h) legend. JI: Jaccard index

For event detection, the segment prediction results were postprocessed to obtain the event prediction results (Fig. 3d). We alternately employed ground-truth event labels and the results of event prediction as references to evaluate the accuracy of event detection. The Jaccard index (JI) [6] was used to evaluate the extent to which a detected event overlapped with a ground-truth event. We defined a JI of ≥0.5 as a correct match to generate a TP event (orange horizontal bars in Fig. 3f and 3g). If a ground-truth event did not have a correctly matched detected event, it was defined as an FN event (yellow horizontal bars in Fig. 3f). If a detected event did not have a correctly matched ground-truth event, it was considered an FP event (black horizontal bars in Fig. 3g). A TN event cannot be defined in the event detection evaluation process. Because the TP events were double counted (orange horizontal bars in Fig. 6f and 6g), we viewed a pair of TP events as a single TP event but independently counted all the FN (Fig. 6f) and FP events (Fig. 6g). This may have led to slightly biased results. Segment detection performance was assessed using F1 scores.

**2.6 Investigation of Noisy Labels and Sound Overlapping**

Herein, we investigated two of the various factors that we indicated in our previous study may influence model performance [9]: noisy labels and sound overlapping. First, we reviewed label quality. Subsequently, we calculated the overlap ratios between I, E, C, and D labels.

# 3  Results

**3.1 Summary of HF_Lung_V1 and HF_Lung_V2 Databases**

Statistics of lung sound files and labels employed in HF_Lung_V1 and HF_Lung_V2 recorded using the Littmann 3200 and HF-Type-1 devices are presented in Table 1. The number of participants increased from 261 in HF_Lung_V1 to 303 in HF_Lung_V2. Moreover, the number of 15-s recordings in HF_Lung_V2 (14138) was 1.45 times that in HF_Lung_V1 (9765), and the total duration increased from 2441.25 min to 3534.5 min. In addition, the number of I, E, C, and D labels increased from 34,095 to 49,610, from 18,349 to 24,581, from 13,883 to 22,535, and from 15,606 to 19,623, respectively.

The number of 15-s files recorded using the Littmann 3200 increased from 4504 to 5156, whereas that of 15-s files recorded using the HF-Type-1 device increased from 5261 to 8982.

**3.2 Model Performance**

F1 scores of segment and event detection are displayed in Table 2. The change trends of the F1 scores derived from expanding from HF_Lung_V1 to HF_Lung_V2 are presented in Fig. 4. The V1 and V2 ticks on the x-axis in Fig. 4a–d represent the number of I, E, C, and D labels in V1_train and V2_train, respectively, and the values on the y-axis represent the F1 scores.

**Table 1** Statistics of lung sound files in the HF_Lung_V1 and HF_Lung_V2 databases.

| Database | | HF_Lung_V1 | | HF_Lung_V2 | | HF_Lung_V1 | HF_Lung_V2 |
|---|---|---|---|---|---|---|---|
| Recording Device | | Littmann 3200 | HF-Type-1 | Littmann 3200 | HF-Type-1 | | |
| Subjects | | 261 | 18 | 303 | 28 | 261 | 303 |
| No. of 15-sec. recordings | | 4504 | 5261 | 5156 | 8982 | 9765 | 14138 |
| Total duration (min) | | 1126 | 1315.25 | 1289 | 2245.50 | 2441.25 | 3534.5 |
| Inhalation | No. | 16535 | 17560 | 18887 | 30723 | 34095 | 49610 |
| | Duration (min) | 257.17 | 271.02 | 294.27 | 494.91 | 528.14 | 789.18 |
| | Mean (sec) | 0.93 | 0.93 | 0.93 | 0.97 | 0.93 | 0.95 |
| Exhalation | No. | 9107 | 9242 | 10480 | 14101 | 18349 | 24581 |
| | Duration (min) | 160.25 | 132.6 | 182.96 | 192.71 | 292.85 | 375.67 |
| | Mean (sec) | 1.06 | 0.86 | 1.05 | 0.82 | 0.96 | 0.92 |
| CASs | No. C/W/S/R | 6984/3974/152/2858 | 6899/4483/534/1882 | 9184/5068/366/3750 | 13351/9048/548/3755 | 13883/8457/686/4740 | 22535/14116/914/7505 |
| | Duration (min) C/W/S/R | 105.90/63.92/1.94/40.04 | 85.26/55.80/7.52/21.94 | 132.93/78.04/5.21/49.68 | 175.01/124.01/7.61/43.39 | 191.16/119.73/9.46/61.98 | 307.94/202.06/12.82/93.06 |
| | Mean (sec) C/W/S/R (sec) | 0.91/0.97/0.76/0.84 | 0.74/0.75/0.85/0.70 | 0.87/0.92/0.85/0.79 | 0.79/0.82/0.83/0.69 | 0.83/0.85/0.83/0.78 | 0.82/0.86/0.84/0.74 |
| DASs | No. | 7266 | 8340 | 7311 | 12312 | 15606 | 19623 |
| | Duration (min) | 111.75 | 55.80 | 112.32 | 169.36 | 230.87 | 281.68 |
| | Mean (sec) | 0.92 | 0.87 | 0.92 | 0.83 | 0.89 | 0.86 |

**Table 2** F1 scores of segment and event detection for inhalation, exhalation, CAS, and DAS detection.

| Training and test datasets | Inhalation | | Exhalation | | CAS | | DAS | |
|---|---|---|---|---|---|---|---|---|
| | Segment Detection | Event Detection | Segment Detection | Event Detection | Segment Detection | Event Detection | Segment Detection | Event Detection |
| V1_Train on V1_Test | 0.806 | 0.840 | 0.624 | 0.637 | 0.527 | 0.438 | 0.712 | 0.596 |
| V2_Train on V1_Test | 0.822 | 0.856 | 0.656 | 0.671 | 0.625 | 0.522 | 0.711 | 0.595 |
| V1_Train on V2_Test | 0.811 | 0.838 | 0.697 | 0.760 | 0.462 | 0.361 | 0.789 | 0.565 |
| V2_Train on V2_Test | 0.825 | 0.851 | 0.705 | 0.766 | 0.558 | 0.434 | 0.786 | 0.562 |

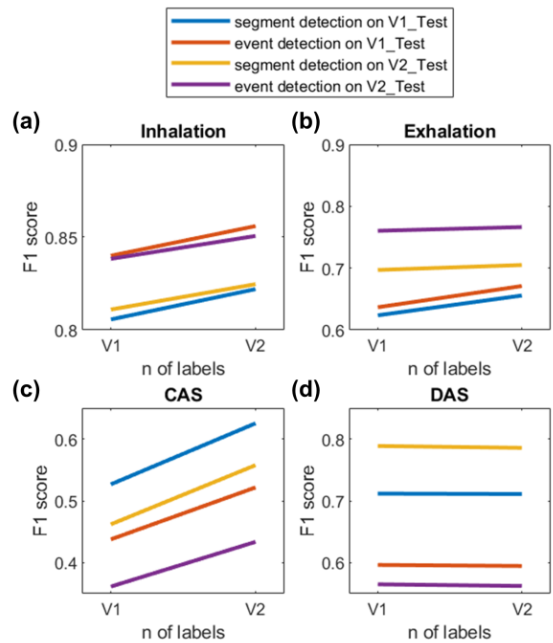

**Fig. 4** F1 scores of segment and event detection of (a) inhalation, (b) exhalation, (c) CASs, and (d) DASs using HF_Lung_V1 (V1) and HF_Lung_V2 (V2)

**3.3 Review of Label Quality**

Fig. 5 illustrates the spectrograms (top graphs in Fig. 5a–e) and labels (bottom graphs in Fig. 5a–e) of five lung sound recordings.

The quality of I and E labels was acceptable. In most cases, inhalations and exhalations can be clearly heard and identified, such as the example displayed in Fig. 5a. However, we can occasionally see wrong labels, such as the exhalation indicated by the white arrow in Fig. 5b, or debatable labels, such as the exhalations indicated by the blue arrows in Fig. 5b. An exhalation was sometimes not identifiable, such as the sound displayed in Fig. 5c.

However, the quality of C labels was unsatisfactory. CASs were often found unlabeled, such as the ones (green arrows) in Fig. 5c. In some cases, the labelers were unsure whether to label a borderline case. For example, the labeler did not label the sound indicated by the first green arrow in Fig. 5d but labeled the sounds indicated by the second and third green arrows although these sounds did not form clear streak patterns on the spectrogram.

As for the quality of D labels, even experts could not arrive at a consensus on the correct labeling of a given DAS. For instance, many noises may have DAS-like patterns and interfered with the labeling process (the purple arrow in Fig. 5e). Furthermore, the DAS patterns in Fig. 5b are sometimes not as clear as those displayed in Fig. 5e.

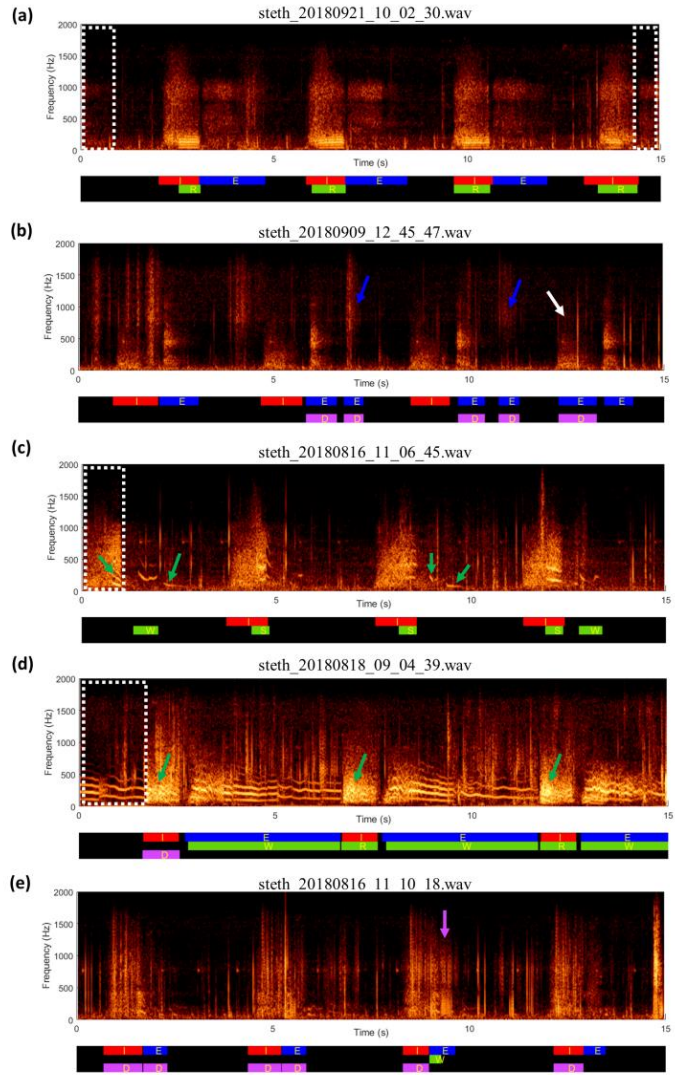

**Fig. 5** Spectrograms (top) and labels (bottom) of five lung sound recordings. In the label graphs at the bottom, the red squares with the letter "I" represent I labels. By the same logic, the blue squares with the letter "E" represent E labels, the purple squares with the letter "D" represent D labels, and the green squares with the letters "S," "W," and "R" represent S, W, and R labels, respectively. The blue arrows in Fig. 5b indicate two sounds with debatable E labels. The white arrow in Fig. 5b indicates an inhalation that was incorrectly labeled as E. The green arrows in Fig. 5c indicate the CASs that were left without a correct S, W, or R label. The green arrows in Fig. 5d indicate the sound recordings that did not form clear streak patterns. The purple arrow indicates some noises that interfered with DAS labeling. The white dashed squares in Fig. 5a, 5c, and 5d indicate the sound events that were left without a label at the beginning and end of a lung sound recording.

### 3.4 Overlap Ratios Between Labels

Table 3 lists the ratios indicating the extent to which a specific type of labels overlapped with the other types of labels in the V2_Train and V2_Test; 94.7% (55.5% + 38.2%) of the duration of D labels overlapped with the I and E labels in V2_Train, and 4.7% (66.1% + 30.6%) of the duration of D labels overlapped with the I and E labels in V2_Test.

**Table 3** Overlap ratios between I, E, C, and D labels in HF_Lung_V2.

| Label | Overlapped with | Overlap ratio (%) | |
|---|---|---|---|
| | | Train | Test |
| I | E | 0.1 | 0.1 |
| | C | 19.0 | 16.7 |
| | D | 18.7 | 28.7 |
| E | I | 0.3 | 0.2 |
| | C | 23.5 | 21.2 |
| | D | 28.1 | 26.6 |
| C | I | 46.5 | 51.7 |
| | E | 27.1 | 32.9 |
| | D | 0.6 | 0.9 |
| D | I | 55.5 | 66.1 |
| | E | 39.2 | 30.6 |
| | C | 0.7 | 0.6 |

### 4 Discussions and Conclusions

In this paper, we reported the effort of expanding HF_Lung_V1 to HF_Lung_V2. The performance of the CNN–BiGRU model improved in inhalation, exhalation, and CAS detection accuracy as the data size increased in HF_Lung_V2. However, we did not observe an improvement in DAS detection, possibly due to the heavy overlapping of D labels with I and E labels.

Small data regions, small power-law regions, and irreducible error regions are present in the curve of the power law of learning [10]. The generalization error (log-scale) decreases as the training set size (log-scale) in the power law region increases [10]. We did not investigate whether the increase in the size of HF_Lung_V2 (which is 1.45 times that of HF_Lung_V1) was in the power law region. The promising improvement in the performance of inhalation, exhalation, and CAS detection encourages us to continue collecting more breathing lung sounds and build a larger dataset.

Notably, the F1 scores of the models trained using V1_Train and tested on V1_Test differed from those reported in our previous study [7] because we retrained the models. Retraining was performed because we conducted event detection evaluation slightly differently. Specifically, unlike in [7], we alternately used ground-truth event labels and the results of event prediction as references. Thus, we may have undesirably counted an FP event and an FN event within the same time period, and the F1 scores of event detection are likely to be lower than those we previously reported.

Label quality review revealed that the quality of I and E labels were just acceptable. However, the quality of C labels was unsatisfactory. This might have occurred because we asked the labelers to perform the labeling mostly on the basis of what they heard and to use the spectrogram only as an aid. We also asked them not to label a sound unless they were sure. Furthermore, some noisy labels may have resulted from mistakes made in operating the labeling software, such as the E label indicated by the white arrow in Fig. 5b. In addition, establishing clear criteria to decide whether a sound can be counted as a CAS or to which types of CAS a sound belongs is difficult. Many borderline rhonchi did not form a clear streak pattern; thus, the labelers were unsure whether to label them (green arrows in Fig. 5d). Upon the exclusion of R labels from the C labels, the model exhibited superior performance in CAS detection (data not shown). As for D labels, the evaluation quality was strongly influenced by the indifferentiable rubbing sound noises. The labelers also tended to ignore the sounds located at the very beginning and end of an audio file because the event was cut off or because the full duration of an event was not heard or observed (white squares in Fig. 5a, 5c, and 5d). The labeling process is currently under reworking such that ground-truth labels can be established. The revised labels will be updated in the future.

DAS detection performance was greatly influenced by sound overlapping. We labeled only the start and end times of a series of DASs (mean duration of each series: 0.86 s; Table 1); we did not label each DAS (mean duration of fine and coarse crackle sounds: approximately 5 and 15 ms, respectively [3]). This caused the overlapping of more than 94% of D labels with the I and E labels (Table 3). Moreover, because we used only the audio recordings containing D labels to train the DAS detection model, the trained model showed tendency to incorrectly identify inhalations and exhalations as DASs. A new dataset that contains more balanced data or a new training strategy is required to solve this problem.

**Declarations**

**Funding**

Taiwan's Raising Children Medical Foundation sponsored the lung sound collection. Heroic Faith Medical Science Co., Ltd. sponsored the data labeling and deep learning model training. This study is supported by Ministry of Science and Technology, Taiwan, R.O.C. under Grant no. MOST 109-EC-17-A-22-I3-0009.

**Conflicts of interest**

Fu-Shun Hsu, Shang-Ran Huang and Yuan-Ren Cheng are employees of Heroic Faith Medical Science Co., Ltd. Chien-Wen Huang and Chun-Chieh Chen are with Avalanche Computing Inc. who are commissioned by Heroic Faith Medical Science to train the AI models.

**Availability of data and material**

HF_Lung_V1 is an open-access lung sound database which can be found at https://gitlab.com/techsupportHF/HF_Lung_V1. However, HF_Lung_V2 is not open to the public.

**Code availability**

Not applicable.

**Ethics approval**

The protocol was approved by the Research Ethics Review Committee of Far Eastern Memorial Hospital (case number: 107052-F). This study was conducted in accordance with the 1964 Helsinki Declaration and its later amendments or comparable ethical standards.

**Consent to participate**

Informed consent is obtained from each research participant whose lung sound were recorded.

**Consent to publish**

The research participants have consented to the submission of their data for academic publication.